\documentclass[draft,showpacs,showkeys,preprintnumbers,amsmath,amssymb, 12pt]{revtex4}

 \usepackage{epsf}

\usepackage{amsmath,amsfonts}

\newcommand{\BC}{{\mathbb C}}

\newcommand{\BR}{{\mathbb R}}

\newcommand{\vev}[1]{\langle #1\rangle}

\begin{document}

\title{The shape of emergent quantum geometry from an  $N=4$ SYM minisuperspace
 approximation}
\author{David Berenstein$\!^\dagger$, Yuichiro Nakada$\!^{\dagger,\ddagger}$ }

 \affiliation{$\!^{\dagger}$Department of
Physics, University of California
at Santa Barbara, CA 93106\\
$\!^{\ddagger}$Institute for Cosmic Ray Research
 University of Tokyo, Kashiwa, Chiba 277-8582 JAPAN}

\begin{abstract}
We study numerically various wave functions in a gauged matrix quantum mechanics of six commuting hermitian $N\times N$ matrices. Our simulations span ranges of $N$ up to 10000.
This system is a truncated and quenched version of $N=4$ SYM that serves as a minisuperspace approximation to the full ${\cal N}=4$ SYM system. This setup encodes aspects of the geometry of the AdS dual in terms of joint eigenvalue distributions for the matrices in the large $N$ limit. We  analyze the problem of determining geometric measurements from these fluctuating distributions at finite $N$ and how fast they approach to the large N limit. We treat this eigenvalue geometry information as a proxy for geometric calculations in quantum gravity in a description where gravity is an emergent phenomenon. Our results show that care is needed in choosing the observables that measure the geometry: different choices of observables give different answers, have different size fluctuations at finite $N$ and they converge at different rates to the large $N$ limit.  
 We find that some natural choices of observables are pathological at finite $N$ for $N$ sufficiently small. Finally, we note that the approach to the large $N$ limit does not seem to follow the expected convergence in powers of $1/N^2$ from planar diagram arguments. Our evidence suggests that different powers of $N$ appear, but convergence to large $N$ is rather slow so the values of $N$ we have explored might be too small to conclude this unambiguously. 
\end{abstract}

\pacs{11.25.Tq, 11.25.Sq ,02.70.Uu, 11.15.Pg}

\keywords{Emergent geometry, AdS/CFT}
\maketitle

\section{Introduction}

The AdS/CFT correspondence has revolutionized how we think about strongly quantum field theories and quantum gravity \cite{M}. Because of these discoveries, classical gravitational techniques can be used to produce non-trivial statements about strongly coupled field theories. On the other hand, the manifest unitarity of the field theory seems to resolve (at least in principle), the information paradox related to the formation and evaporation of black holes \cite{Hawking}.

Obviously, if the two sides are truly equivalent, we should ask that the quantum gravitational theory encode individual quanta and other phenomena that we associate to weakly coupled field theories. By the same token, we should be able to answer geometric questions posed in the gravity theory by studying the degrees of freedom of the quantum field theory directly. The usual AdS/CFT dictionary \cite{GKP,W} usually works as a black box: it permits us to compare calculations between the AdS and CFT setups, but it does not explain how the degrees of freedom of the field theory reorganize themselves as geometry. Nor does it tell us how geometry breaks down. What is needed is a theory of emergent geometry: a reason why higher dimensional geometry is the right answer for certain classes of questions. Solving the AdS/CFT correspondence requires doing this. The full answer to this question is still beyond what is currently known. However, a model that seem to capture some of these features is available \cite{BLargeN} and can be used to explore the passage from a microscopic description of the degrees of freedom to collective degrees of freedom that represent geometry. 

Given a model of emergent geometry, one can then ask for the simplest geometric questions, like what is the size of an object or geometric feature. This will be the main purpose of the present paper: to understand and explore how one can measure geometric properties  of the dual spacetime directly in the quantum field theory. We will do this by Monte-Carlo simulations of a collection of minisuperspace wave functions in field theory, following the ideas presented in \cite{BCotta,BCL}. Numerical methods to study gravity problems with a dual quantum mechanics have been advocated also in \cite{JW,CW,AHNT,CW2,HMNT,HHNT,CW3}. In particular these other studies try to match the black hole specific heat and perhaps even the radius of the horizon for a black hole made of D0 branes in ten dimensions. Those simulations use the full $N\times N$ matrix degrees of freedom in a Euclidean time formulation, so taking $N$ large is problematic. Other approaches to study $N=4$ SYM for large $N$ numerically use clever truncations to matrix quantum mechanics \cite{IIST,IKNT}, but they suffer from similar limits on taking $N$ large. 

The main problem at hand is that in order to compare gravity and the field theory, in the regime where gravity is described by a semiclassical expansion, we are required to solve a field theory at strong coupling. This is extremely hard in general and this is why numerical methods might offer a good insight into the strong coupling behavior of these field theories.
However, as is well known, the gravitational limit also requires solving the field theory at large $N$. Current lattice methods 
can in principle solve four dimensional theories for moderate values oF $N \sim 10$, but taking $N$ large is too difficult.
Remember that in the $AdS/CFT$ correspondence the radius of the five sphere and also of the $AdS$ sphere in Planck units scales like $N^{1/4}$. Thus, in order to have a resolution of the sphere which is many times the Planck length, let us say by an order of magnitude, we require four orders of magnitude on $N$. Such numbers are beyond what can be done with a full simulation. Clearly, some approximation to the full field theory problem is required that reduces effectively the number of degrees of freedom to something manageable. Only at this stage simulations become a viable option to analyze the problem. 

Progress in this direction was made in \cite{BLargeN}. It was argued there that at strong coupling, when the ${\cal N}=4 $ SYM field theory is compactified on a three-sphere,  the configurations of commuting (constant) scalar matrices dominate the infrared dynamics. Toy models that exhibit this property were exhibited in \cite{BHH}, giving additional indirect evidence for this proposal in the case of ${\cal N}=4 $ SYM. 

The nature of the argument that leads to this conclusion requires various steps. First one argues that such classes of configurations are required for describing some supersymmetric states, so the truncation to such variables might be better behaved than naively expected. 
Afterwards one argues that the quartic scalar potential of the theory should be minimized. This second step is accomplished by systems of commuting matrices. The dominance of this configurations dynamically quenches the
other off-diagonal degrees of freedom, meaning that they become heavy and their contributions can be ignored to some degree.

The end product is a model describing only a handful of the degrees of freedom of the full ${\cal N}=4$ SYM theory, and since the other degrees of freedom are essentially ignored, we can call this setup a quenched  minisuperspace approximation. We will use the name quenched in the rest of the paper. 
What has been found is that the distribution of eigenvalues of the commuting matrices (in this approximation) has a geometry of its own \cite{BLargeN, BCV}. The matrix eigenvalues experience some quantum induced repulsion (similar to the VanderMonde repulsion of eigenvalues in matrix models \cite{BIPZ}), arising from the volume of the gauge orbit of the configurations, and a confining potential due to the coupling to the curvature of the sphere (this is realized as an effective mass term for the eigenvalues). Repulsion wins at short distances, and the confining potential wins at long distances. Because of these competing effects one expects an equilibrium at finite radius and shape. 

 In the large $N$ thermodynamic limit the eigenvalues can form a geometric distribution that depends on the particular wave function of the system that is chosen. For the ground state, the wave function is known \cite{BLargeN}. We will call this wave function $\psi_0$. For this wave function the eigenvalue distribution is an $S^5$ submanifold of $\BR^6$, where the radius is of order $\sqrt N$ in field theory units. Locality in $S^5$ is induced from locality in $\BR^6$: the masses of off-diagonal degrees of freedom depend on the distances between eigenvalues in $\BR^6$.  Modes connecting nearby eigenvalues correspond to lighter degrees of freedom, whereas modes connecting distant eigenvalues correspond to very heavy degrees of freedom that decouple sooner. Hence the distance in $\BR^6$ is meaningfully describing a notion of locality: how coupled the different degrees of freedom are to each other when integrating out heavy fields.
 
One can argue this way that this distribution describes the $S^5$ of the $AdS_5\times S^5$ dual geometry. Other supersymmetric wave functions are obtained by multiplying the ground state wave function with holomorphic polynomials of the variables, properly symmetrized between the eigenvalues. 
This type of argument has also been generalized to orbifolds \cite{BCorr,BCott} and conifolds \cite{Bcon,Bhart}, giving a large number of qualitative and quantitative tests of this idea. However, reproducing the exact geometry of supergravity from the field theory is still beyond the current understanding of the AdS/CFT dictionary. A particularly difficult question to answer is how different field theory degrees of freedom are encoded and localize in different parts of the AdS geometry, how the holographic bound is realized in practice \cite{SW} and in particular how to relate these questions to the holographic renormalization group \cite{BK,BVV} (for a recent review see \cite{HRG}).

The upshot is that some geometric information can be extracted from these wave functions, but the said geometry contains only partial information about the full ten dimensional geometry that supergravity would see. Since to date there is no convincing way to extract the ten dimensional geometry directly from the field theory without assuming a correspondence, we will use the limited  geometric information that can be derived from the field theory in these approximations as a proxy for the full dual (quantum) geometry. Naturally, one can argue that this is the same problem of comparing apples to oranges and therefore the results are of limited value. Although this is technically true, a toy model that performs this reorganization of degrees of freedom into some type of geometry is extremely valuable because some of the conclusions might be universal enough that they apply even in the general case. A case in point is the usage of ${\cal N}=4 $ SYM at finite temperature as a proxy for finite temperature QCD, and the use of the gravity dual to SYM to make predictions about the viscosity of the quark gluon plasma \cite{PSS}.

Contrary to other approaches of quantum gravity that attempt to build a spacetime from nothing: like those growing spacetime from a point in the no-boundary proposal for the wave function of the universe \cite{HH}, or the causal set approach to quantum gravity \cite{Henson}; in this setup there is no unique preferred wave function, and different wave functions correspond to different geometries. Some wave functions might not correspond to a single geometry either: superposition of geometries is possible. 
These states would be like a Schr\"odinger cat state: a simple measurement would destroy the superposition.
However, when we can barely understand a single geometry at this stage of the research program, studying this feature of superpositions of geometry is not desirable.
The setup we are describing is similar to string states as solutions of the string equations of motion in a background spacetime: for each vertex operator of dimension $(1,1)$ (this is how we describe quantum states via the operator-state correspondence) one can  associate a different motion to the string in the semiclassical setup, and that motion defines a geometric trajectory for the string. Neither of these motions is preferred a priori, except as described by some preparation of an experiment. What we are studying here is the same: how different wave functions in our quenched minisuperspace description correspond to different geometries. Our ultimate goal is to establish a tractable dictionary between geometries and wave functions in this very limited context and to study how the fluctuations inherent to the wave function give fluctuations of geometric data. Our aim is not to be exhaustive, but to instead develop some intuition about emergent geometry with some simple wave functions that are easy to evaluate numerically. 

It has also been found that by changing the wave function, the geometry of the field theory degrees of freedom  admits different topologies in the thermodynamic limit. Numerically this has been observed in \cite{BCotta}. This is also what is expected from the classification of half BPS states in
the dual supergravity theory \cite{LLM}, where topology changing transitions are allowed. The simplest wave functions which do this are powers of determinants of one of the complex matrices $Z$ (these are wave functions describing many giant gravitons on top of each other, at weak coupling they are described by similar formulae involving operators of the full matrices \cite{BBNS,CJR}. At strong coupling, the giant gravitons condense into a topology change once the gravitational backreaction is taken into account). The matrix $Z$ is a diagonal matrix. The eigenvalues represent one of the three complex coordinates determining $\BR^6\simeq \BC^3$ as a complex manifold. Such wave functions are of the form
\begin{equation}
\psi_Q= \det(Z)^Q \psi_0\label{eq:wf1}
\end{equation}
where $Z+ X^1+iX^2$ for example, and $Q$ is an arbitrary integer.

When $N$ is finite, the system (with the wave function) should be considered as a version of quantum geometry at finite values of $\hbar$. The quantity $1/N$ plays such a role in the 
$AdS$ geometry. After all $N^{1/4}$ measures the size of $AdS$ in Planck units. Thus, finite $N$ effects and how fluctuations depend on $N$ can be said to encode quantum gravity corrections.

The idea of this paper is to gain some understanding of emergent quantum geometry. Since we do not have direct access to the ten dimensional geometry from current methods, we will use the quenched field theory geometry to explore the problem of convergence to large $N$ and the problem of determining the shape of the geometry at finite $N$. A similar problem has been studied in \cite{Hashimoto, AHHS} for the case of fuzzy geometries of brane embeddings (where the brane configurations are non-abelian). In our case the notion of shape needs to be refined with respect to the thermodynamic limit. What we call geometry according to the prescription \cite{BLargeN,BCV} is characterized by the (support of the) large $N$ density of eigenvalue distributions, properly rescaled by the size of the ground state, $\sqrt N$. This can be shown to be a singular distribution of $\BR^6$ in the thermodynamic limit, namely the eigenvalues form a manifold of lower dimension, for some general class of  BPS wave functions \cite{BLargeN, Bhart}. However, this is just the saddle point description at large $N$ of the wave function. This procedure does not take into account that the eigenvalues are discrete, or that there are fluctuations around this saddle point.

At finite $N$ we expect that there are corrections to this geometric saddle point approximation of the wave function. These quantum corrections will be due both to granularity of the constituents (the system has finitely many eigenvalues in $\BR^6$), as well as due to the quantum spread of the wave function (so that the support of the wave function is not singular). The first one is non-perturbative in nature, as it provides a cutoff on the number of independent degrees of freedom characterizing the geometry. It also suggests that the system might have some 'granular' features. The second class of corrections can be used to represent the ordinary quantum fluctuations of the geometry. 

Given a wave function for these degrees of freedom, we would ask the following simple questions:
\begin{enumerate}
\item What is the topology of the distribution?
\item What are the geometric measurements of the holes of the geometry?
\item How big are the quantum fluctuations?
\end{enumerate}

Notice that when we consider $N$ finite, the notion of topology and shape becomes fuzzy, due to both the fluctuations of the eigenvalues, as well as because of the discreteness of the eigenvalues themselves. First, the eigenvalues have some probability to be in any distribution, except on sets of measure zero. Thus the support of the distribution is not a very useful notion of geometry, but maybe it is enough to describe the topology. However, if one is trying to find sets of measure zero numerically this might not be easy to extract from a simulation in the general case. Secondly, measurements on fluctuating object shapes will have some inherent randomness and there will be errors in measurements. If the particles can be in any spatial distribution, there are no minimal distances to define sizes either. Thus how one chooses to sample the distribution to define a notion of size becomes important as well. 

In such a situation, we have to change the questions to something that makes sense computationally. We need observables such that in the thermodynamic limit ($N\to \infty$) they recover the classical measurement problem with probability arbitrarily close to one at finite sampling (taking a single measurement on a typical configuration, within some prescribed error bars). Some observables are better suited to this task than others. Moreover, different observables will differ at finite $N$ by $1/N$ corrections (see for example the discussion in \cite{KT} for fuzzy sphere membranes). So the real question we need to ask is what are the optimal choices of observables that characterize geometric measurements and how fast do they approach the thermodynamic large N limit?

As we said above, we will approach this problem by doing numerical simulations.  The basic setup for these simulations was described in \cite{BCott} and \cite{BCL}. The Monte-Carlo model we use simulates the probability distribution of eigenvalues from the square of the quenched wave function (including the measure effects). This is very different in spirit than lattice approaches. It also is free from sign or phase problems. Also, because our matrices are diagonal, we have only $6N$ degrees of freedom, so $N$ can be made large at moderate cost. Using the same Monte-Carlo code, we will address the questions above for a particularly simple class of wave functions, namely those described by equation \ref{eq:wf1}. For these wave functions the geometry 
of the eigenvalues looks like a five-sphere with a hole in the center, so we are not studying the process of topology changes. We are fixing the topology at the beginning. The issue we will study seems very mundane: how do we measure the 
radius of the hole?
Since these are wave functions of fixed topology, we will not address the first question above at all: the topology is known so we do not need an algorithm to figure it out. We also have chosen the orientation of the feature on the sphere and do not have to orient the data to analyze the geometry.  
Given this, we can see by eye the topology on these simple situations by projecting the particles positions on the $Z$ plane. The more general problem of also determining the topology for a random configuration with a complicated wave function would be hard: we would need to find a way to do pattern recognition on these distributions to define topology. Regarding measurement of size, we propose various definitions of the size of the simplest geometric features  and we study their virtues and failings. In particular, to simplify matters further, we want to characterize the simplest such non-trivial feature: the radius of the hole compared to the radius of the sphere. In the end, from the different choices of definitions of the radius at finite $N$, one such class of measurements seems to give an optimal solution to the problem. For this particular class of observables we can then address the $1/N$ corrections in more detail.  The definitions we use are simple to describe, but that does not mean that they are optimal. We can only say that they are optimal only within the choices we have.

The paper is organized as follows. In section \ref{sec:wf} we describe some basic aspects of the wave functions and statistical distributions that we simulate later on. In particular, we show how the thermodynamic limit should be taken: $N\to \infty$, keeping $q=Q/N$ fixed.  We also pay special attention to how a factor of $N^2$ appears in front of the energy for the thermodynamic limit after appropriate rescalings. We also describe various functions that allow us to define the size of the 
features of the geometry of the eigenvalue distributions. The definition of the size depends on taking a limit $k\to \infty$ of such measurements, so that one expects a well behaved $1/N$ expansion at fixed $k$.
  We also review the argument that leads to a singular distribution of eigenvalues in the saddle point approximations. In  section \ref{sec:nr} we describe our numerical results for the wave functions. We describe data sets in various types of analyses. First we describe individual data sets for fixed $Q, N$. This lets us see the topology of the configuration from the plots and we also see that the support of the distribution of eigenvalues if of codimension one at large $N$, with fluctuations. We can also compare the different characterizations of the radius of the hole to see how big the fluctuations are on the various measurements and why some statistical characterizations of this feature are pathological (they are dominated by fluctuations). For the measurements that are well behaved we find that they converge at different rates, that one can mathematically prove some inequalities between them and that one such type of measurement converges faster and is optimal given our choices. Given this information, we then study the problem of varying $N$ at fixed $q=Q/N$, to see how the large $N$ limit is approached. Our numerical results show that the convergence to large $N$ is slower than would be expected based on large $N$ counting arguments. Finally, we also study the radius of the hole by varying $q$ at fixed large $N$ ($N=10000$) and obtain a scaling exponent of the radius as a function of $q$ for small $q$. These results are different than those that were found previously in \cite{BCotta}, that were obtained for much smaller values of $N$. We then conclude.

\section{Wave functions and Geometric observables}\label{sec:wf}

The matrix model of commuting hermitian matrices  $X^a$ \cite{BLargeN} leads (by diagonalization of the matrices) to a model of $N$ particles in six dimensions whose quantum  Hamiltonian
is given by
\begin{equation}
H =\sum_i -\frac 12  \frac 1 {\mu^2} \vec \nabla_i \mu^2 \nabla_i +\frac 12 |\vec x_i|^2 
\end{equation}
where $\vec x_i$ are the positions of the particles in $\BR^6$ and $\mu^2= \prod_{i<j}|\vec x_i -\vec x_j|^2$ is a generalized Vandermonde determinant. The $x^a_i$ are the $i$-th eigenvalue of the matrix $X^a$. The six $x$ variables per eigenvalue can be written in complex notation in terms of three complex numbers  $ \sqrt 2 z^a =   x^{2a-1}+ i x^{2a}$, which can be arranges as a vector $\vec z$ and their complex conjugates $\vec {\bar z}$ per each eigenvalue. 

The ground state wave function is given by \cite{BLargeN}
\begin{equation}
\psi_0(x) = \exp\left(-\frac 12\sum_i |\vec x_i|^2\right) = \exp(-\sum_i \vec z_i \vec{\bar z}_i)
\end{equation}
The associated probability density for the particle positions is then given by
\begin{equation}
p = |\psi_0|^2 \mu^2 =\exp\left(-\sum_i |\vec x_i|^2 +  \sum_{i<j} \log| \vec x_i -\vec x_j|^2 \right)
\end{equation}

It is also easy to show also that the wave function above is an eigenfunction of the following similar looking Hamiltonian
\begin{equation}
 {\mathfrak H} = \sum_i -\frac 1 {\mu^2} \nabla_{z,i} \mu^2 \nabla_{\bar z,i} + |\vec z_i|^2
\end{equation}
with the derivatives being holomorphic and anti-holomorphic in the prescribed order. To prove this, one has to use the fact that $\mu^2$ is a scaling function under rescalings of the $z$ variables that leave $\bar z$ fixed (we are treating them as independent variables, rather than complex conjugate variables for that). The complete argument, which is beyond the scope of the present paper, is very similar to the one appearing in \cite{BLargeN,Bhart}.

The next set of simplest wave functions in this quenched dynamics are those associated to BPS states. They are described by holomorphic wave functions of the $z$ variables multiplying the ground state wave function and these are symmetrized over the eigenvalues (this is is a discrete gauge symmetry of permutation of eigenvalues). These are not exact wave functions of $H$, but they are argued to be close to exact wave functions of the system.
However, it turns out that all the homogeneous holomorphic wave functions are eigenfunctions of $\tilde H$, with energy $E-E_0$ equal to the degree of the polynomial
of the $z$. As such, the Hamiltonian $\tilde H$ seems more appropriate to capture all BPS states exactly than $H$, since the holomorphic wave functions do not seem to be eigenfunctions of $H$, except under certain approximations in the large N limit \cite{Bhart}. In general, the set of BPS solutions of the classical equations of motion on $S^3$ are in one to one correspondence with the moduli space of vacua of superconformal field theories. The quantization of these configurations is a particular holomorphic quantization of the moduli space of vacua \cite{Bcon}, which requires a measure. The Hamiltonian $\tilde H$ would give the exact answer for the chiral ring state energies and has the correct measure of the ground state wave function in the classical approximation to these solutions, which is the same quenched approximation we have been studying. This is the main reason to introduce it above. However, nothing else that we do depends on that fact.

The simplest wave functions that change the topology are those that involve many (maximal) giant gravitons on top of each other. The giant gravitons in the type IIB gravity theory in $AdS_5\times S^5$ are extended brane configuration with the same quantum numbers as gravitons \cite{MST} that wrap an $S^3$ inside $S^5$. If one places many of them on top of each other one expects a throat to form in the geometry near the location of the branes and to excise the branes from the geometry and replace them by fluxes \cite{LLM, KS, Vafa}. The new topological feature is associated to the removal of the branes and the capping of the region of spacetime where they have been removed.

In the dual field theory  these are described at weak coupling in \cite{BBNS} and can be related to a system of free fermions on a two-dimensional phase space \cite{CJR,Btoy}. This picture, in terms of incompressible quantum droplets is recovered geometrically in the half-BPS geometries \cite{LLM}, where the exact geometries of the capping are understood for half BPS D-branes with arbitrary shapes. These giant graviton states have been extensively studied at weak coupling. 
A recent review of many of these results can be found in \cite{KochM}.

Now, just having the map of droplets to states and to gravity does not mean that we understand the ten dimensional geometry from field theory. It just gives us a correspondence of particular solutions, and our task would be to explain the ten dimensional geometry from this data: how did the extra dimensions arise and why is there locality, time bending, etc., in the geometries. This is too hard with present techniques, which is why we will look at the field theory probability distributions for the quenched wave functions directly.

The wave functions we consider are given by
\begin{equation}
\psi_Q= \det(Z^1)^Q \psi_0
\end{equation}
where $Z^1= X^1+i X^2$ in terms of the matrices and $Q$ is an integer determining the number of giant gravitons. These are also given by $z_i^1$ coordinates. Notice that $Q$ is an integer in order for the wave function to be single valued. The normalization of $\psi$ does not matter to determine the eigenvalue distribution, after all, wave functions are rays in a Hilbert space.

The associated probability distribution of the eigenvalues (particles in six dimensions) is given by
\begin{equation}
p \sim \exp\left( - \sum_i |\vec x_i|^2 +  Q \sum_i \log(|z^1_i|^2)+ \sum_{i<j} \log|\vec x_i -\vec x_j|^2\right) \label{eq:pdist}
\end{equation}
Notice that this probability distribution can be interpreted as a Boltzmann gas of particles with two body repulsive logarithmic  interactions in 6 dimensions, subject to a background potential. The logarithmic repulsions in six dimensions are long range. Because of that, the associated statistical mechanical problem is not standard. If $Q=0$ the background potential is confining towards the origin. If $Q\neq 0$, the background potential is confining in generic directions, but it is repulsive from the $z^1=0$ locus with a logarithmic repulsion (this is, particles are repelled from the locus $x^1=x^2=0$). 

In the gravity theory, we usually take the large $N$ limit to go to the classical limit. In this case, we can take the large $N$
limit of the Boltzmann gas which is a thermodynamic limit of large number of particles. In the thermodynamic limit we should be able to replace individual eigenvalues by a density of eigenvalues on $\BR^6$, given by $\rho(\vec x)$. The probability density for the eigenvalue distribution is then given by
\begin{equation}
p (\rho) \propto \exp\left(
- \int d^6x \rho(x) \left[ \vec x^2+ Q \log(x_1^2+x_2^2)\right]+\frac 12 \iint d^6x d^6y \rho(x)\rho(y)\log|\vec x-\vec y|^2\right)\end{equation}
The density is constrained so that $\rho\geq 0$ and $\int d^6 x \rho(x) =N$. As shown in \cite{BCV}, the size of the distribution for $Q=0$ is of order $\sqrt N$. So we can rescale the coordinates $\vec x = \sqrt N \vec a$ and the density
$\hat\rho(a) d^6 a =  \rho(x)N^{-1} d^6 x $. In this way, we have that the normalized distribution $\hat \rho$ satisfies
\begin{equation}
\int \hat \rho(a) d^6 a = 1\label{eq:constraint}
\end{equation}
If we do this substitution we find that
\begin{equation}
p(\hat\rho) \propto \exp\left(
- N^2 \int d^6a \hat\rho(a) \left[ -\vec a^2+ q \log(a_1^2+a_2^2)\right]+\frac {N^2}2 \iint d^6a d^6b \hat\rho(a)\hat\rho(b)\log|\vec a-\vec b|^2\right)
\end{equation}
where $q=Q/N$. Notice that the extra numerical factors that are obtained from the rescaling of $x$ inside the logarithms give a constant, due to the constraint $\int \hat \rho =1$.  This can be absorbed in the normalization of $p$. In this new setup, we see that a limit should be smooth at large $N$ if $q$ is held fixed in the process. Also, the factor of $N^2$ in front of the exponential term shows that when $N\to\infty$ we should be able to use a saddle point configuration for $\hat \rho$ with confidence. It is also clear that $1/N^2$ plays a similar role to $\hbar$ in semiclassical physics, where the saddle point equations to a Euclidean path integral involve finding the extrema of $\exp(-S/\hbar)$. $1/N^2$ can also be interpreted as a temperature, so large $N$ cools the system and suppresses fluctuations.
 This is exactly what is expected from the match to supergravity \cite{M} and to the planar diagram counting \cite{'tH}.

The equations of the saddle point are the following
\begin{equation}
C- \vec a^2 + q  \log(a_1^2+a_2^2) + \int d^6 b \hat\rho(b)\log|\vec a-\vec b|^2=0 \label{eq:saddle}
\end{equation}
where $C$ is a Lagrange multiplier enforcing the constraint (\ref{eq:constraint}). These equations are valid on the support of $\rho$.  We can prove that these equations lead to a singular support for $\rho$. We do this by contradiction.

Assume that a smooth locus for the support of $\hat \rho $ is allowed. In that case, we can differentiate the equation \ref{eq:saddle} six times, with the Laplacian in $6$ dimensions to the cubed power. We find then that
\begin{equation}
(\nabla^2)^3 q  \log(a_1^2+a_2^2) +\int db^6 \hat\rho(b) \nabla^2 \log|\vec a-\vec b|^2=0
\end{equation}
as the first two terms obviously are cancelled. Notice that 
\begin{equation}
\log (a_1^1+a_2^2) \simeq \log(z^1)+\log(\bar z^1)
\end{equation}
so that away from $z^1=0$ we have that 
\begin{equation}
\nabla^2\log (a_1^1+a_2^2)=\nabla^2\left( \log(z^1)+\log(\bar z^1) \right)=0
\end{equation}
because the function is harmonic: a sum of a holomorphic plus an antiholomorphic piece.
Also, is is easy to show that $(\nabla^2)^2 \log(\vec a -\vec b) \propto \delta^6(\vec a-\vec b)$. With these, we would find that $\hat\rho= 0$. This argument contradicts that $\int \hat\rho=1$, so the assumption that $\hat \rho$ is smooth does not work. This argument has been given before in \cite{BLargeN, Bhart}, and we are presenting it here for completeness.

What this indicates is that the density distribution of particles is singular.  Since for $q=0$ the original probability distribution is spherically symmetric, it is natural to assume that the associated distribution is spherically symmetric. The radius was computed in \cite{BCV}.  This has also been checked numerically in \cite{BCotta, BCL}. A generalization to all Calabi-Yau cone conformal field theory singularities for this result can be found in \cite{Bhart}, where it was seen that this procedure matches various other supergravity calculations in the absence of symmetry.

For $q > 0$ but very small, we expect that the distribution does not change much, except in the region $z^1\simeq 0$.
Thus we expect that the distribution gives a five dimensional manifold of $\BR^6$, possibly with a boundary, and that the region near $z^1=0$ has no density of eigenvalues present. After all, the effective potential repels all particles from the locus $z^1=0$. This indeed was seen numerically in \cite{BCotta}, but the treatment was not systematic. This paper 
studies this issue in detail.

In this situation, we see from the saddle point equations that the radius of the hole should be a function of $q$ in the large $N$ limit. However, for $N$ finite there are corrections. The corrections from fluctuations about the saddle point are of order $1/N^2$, because $N^2$ is playing the role of $\hbar$ in the saddle point approximation. There are also corrections due to granularity, for which there is currently no
theoretical understanding as to how they systematically affect the correlations.

The projection of the density of particles in the $12$ plane gives us a disk with a hole in it. Let us assume that this picture from these simulations is correct geometrically. Therefore in the large $N$ limit the saddle point density has no fluctuations (they are suppressed by $1/N^2$). Because of this we should be able to define a geometric radius of the inner radius of the distribution at $N\to \infty$. This should be measurable.

Now, we need to find a good geometric description of  the size of the hole that can be evaluated numerically for configurations that are not strictly at $N=\infty$, but that have some fluctuations. This is, we want to find a radius function
of the distribution that we can average over configurations
\begin{equation}
r_0(N,q)  =\vev{ r_0(\hat \rho, N)}
\end{equation}

We want the function $r_0(\hat \rho,N)$  to be as well behaved as possible. This is, we want the fluctuations of $r_0$ to be small. There is more than one such function. Let us consider various options.

The arguments of planar diagrams suggest that good functions are traces over matrices. These are given in the thermodynamic limit by integrals over $\rho$ (equivalently, integrals over $\hat \rho$). Consider for example the following numbers
\begin{equation}
f_k= r_k^{-2k} =  \int \hat \rho(a) \frac 1{|(a_1)^2+(a_2)^2|^{2k}} \label{eq:radiusdef}
\end{equation}
The quantities $r_k$ are natural. They weigh the points in the distribution by how far they are from the $z^1=0$ locus.
For very large $k$, these are dominates by the smallest value $r_0$ that $\hat \rho$ can attain. These are also easy to compute numerically. This was the rationale for using them in our numerical codes and this choice was made before taking the data. We collected these for various $k$ for many statistically independent configurations at different values of $Q,N$.
We will explain the data later.

Thus, we have our first definition of the radius $r_0$. This is given by
\begin{equation}
r^{(s)}_0= \lim_{k\to \infty}\vev{r_k^{-2k}}^{-\frac1{2k}}  = \lim_{k\to \infty} \vev{f_k}^{\frac{-1}{2k}}
\end{equation}
Notice that this is an improvement over the definition of the radius given in \cite{BCotta}, where the average of the closest five points to the $z^1=0$ locus were used. That definition does not seem to be well behaved at large $N$. Also, we are averaging a trace, so we can expect that the fluctuations are of order $1/N^2$ (this is a typical planar counting argument and at this stage it is just a theoretical intuition to set the size of errors). 
Here we have a definition that is more democratic in the particles. The $(s)$ superscript is a reminder that this is a simple definition. However, this definition runs into other problems that are not immediately obvious. This is that the average of the expression (\ref{eq:radiusdef}) can be infinite, giving us $r^{(s)}=0$. This is because we are taking a double limit $N\to \infty$, $Q/N$ fixed $k\to \infty$, but there can be a problem with the order of limits, and indeed there is. Notice that when $k>Q$ the distribution for a single eigenvalue has a zero at $z^1=0$ in $|z^1|$ of order $2Q$. But that the trace is given by $\sum_i 1/|z_i^1|^{2k}$, which has a pole of order $2k$ in the corresponding variable. If $k>Q$ the integral of the probability distribution over the position of the particle diverges.
However, if we take $Q/N$ fixed, and $k<Q$, we have a finite answer. Thus the limit radius $k\to \infty$ would give $0$, which is not right. This means that in this case we are not allowed to take the limit $k\to \infty$. In this situation, we would say the observable is dominated by large fluctuations. We clearly want to avoid these classes of observables to define the shape.

The next simplest expression is given by changing the place where we do the averaging in the above
\begin{equation}
r^{(r)}_0 = \lim_{k\to \infty} \vev{r_k}
\end{equation}
This is, we average a root of a trace, as opposed to take the root of the average of a trace. One can check explicitly that the problem with the poles in the trace dominating the expression disappears now. We are taking the root before averaging, so near the bad locus the behavior of the function $r_k$ is not badly behaved, so the system is not heavily weighed towards these configurations.
Now we can  now take $k\to \infty$ safely. The $(r)$ symbol notation is a reminder to indicate that we are taking a root.
If we take $k\to \infty$ in a finite sum, one can show that the sum is dominated by the single eigenvalue closest to the $z^1=0$ locus, so that recovers a definition close in spirit to the one in \cite{BCotta}. It should be emphasized that
we are not required to take $k\to \infty$. Each of these gives a definition of a radius of the configuration, so we can study the result for various $k$ and collect them together. We should see convergence to a limit in the numerical exploration of these configurations. For our setups the $k$ will vary between one and fourteen. If the inner radius is at $1/2$ of the full 
size of the geometry, the suppression of the outer edge contribution relative to the inner edge in a given sum can be as large  as $(0.5)^{28}\simeq 10^{-9}$ for the case of maximum $k$, but for particles sitting at $r\simeq 1.1 r_{in}$ the suppression is only a factor of $14$ smaller, so they should still be contributing. Of course, if the inner radius is very small, the near inner radius region is also very small and the suppression of the other regions will be more pronounced.

There is a third definition that is useful, and that is the following
\begin{equation}
R_0 = \lim_{k\to\infty} \vev{\left(\frac{r_{k+1}^{2k+2}}{r_k^{2k}}\right)^{1/2}}=\lim_{k\to\infty}\vev{\left(\frac {f_{k}}{f_{k+1}}\right)^{1/2}}
\end{equation}
This is another way of removing the problem with the poles, and instead of a root of the single trace expressions, we are taking a ratio of these expressions for different $k$. In the ratio, if the numerator gets big when one particle goes towards the $z^1$ locus, the denominator also gets large at the same time, and in the observables the average is not unduly dominated by large fluctuations.

It turns out that this last definition, from the ones we have given, is optimal. We will explain how this works in the next section, where we can see directly in the data why this is so. Notice  moreover, that if we take $N\to \infty$ first, one can use large $N$ factorization (one expects small corrections in inverse powers of $N$), so in the limit we also have that
\begin{equation}
R_0^2|_{N\to\infty} = \lim_{k\to \infty}\vev{\frac {f_{k}}{f_{k+1}}}=\lim_{k\to \infty}\frac {\vev {f_{k}} }  {\vev{f_{k+1}}}
\end{equation}
so it is clear that the different definitions are not too different from each other at large $N$. However, they behave very differently at finite $N$.

All of these functions are chosen because when we take $N\to \infty$ at $q$ fixed, they have (at least in principle) a theoretically controlled approach to large $N$ that can be evaluated around the saddle point configuration. There are other reasons to choose inverse powers of the radius: in the paper of Chen et al. \cite{CCS} it was shown that inverse powers of these matrices also appear in the description of closed string states in the presence of backgrounds with topology changes. Also, the moments of the eigenvalue density distribution serve as a characteristic description of shape that determines the half BPS states, and these can also be measured gravitationally in the AdS dual geometry \cite{BCLS} ( these particular calculations use the half BPS states in the free fermion realization of weak coupling, and only describe the ten dimensional geometry indirectly by comparing to the LLM solutions \cite{LLM}).

Since in our case the density does not go all the way to the zero region $z^1=0$, the  moments of the distribution with inverse powers of the radius can also be taken (which is the definition we used). These are hard to read from the asymptotic solution in gravity, but we are not forbidden from doing that in the field theory geometry observables.

\section{Numerical Results}\label{sec:nr}

The probability distributions described by equation (\ref{eq:pdist}) are easy to simulate. The algorithm used to calculate these configurations was described in detail  in \cite{BCotta}. It operates via a standard metropolis algorithm, updating one particle at a time in the Boltzmann gas. The initial configuration is distributed randomly on a cube. The simulation waits for relaxation before taking data. The only new innovation for the current project is to record the sums of the inverse powers of the radius in the $12$ plane of the configuration, as required by the discretized version of  equation (\ref{eq:radiusdef}). 

\subsection{Data sets at fixed $Q,N$}

\begin{figure}[ht]
\epsfxsize=5 cm \epsfbox{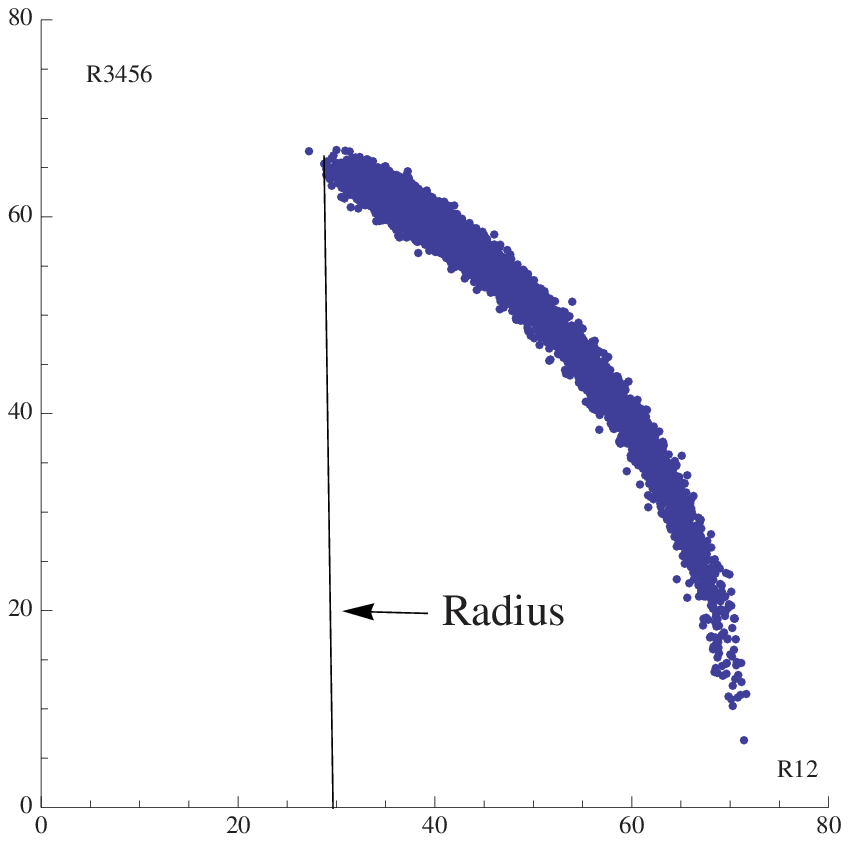}\ \ \epsfxsize=7 cm \epsfbox{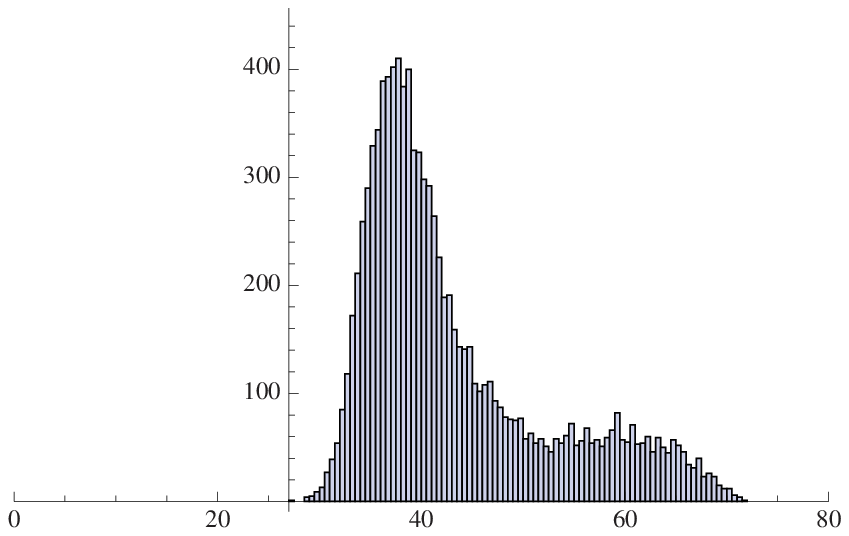}
\caption{The distribution of radii of the configuration. The plot on the left shows that $r_{3456}$ can be considered as a function of $r_{12}$, as expected by the singular nature of the saddle point distribution. It also shows that the radius of the distribution makes sense, except for one outlying point that one can call a statistical fluctuation. The plot on the right bins the number of particles in $r_{12}$ values in increments of a half. One notices a strong peak in the distribution, but it does not happen at the inner edge of the distribution. The sample is taken with $N=1000, Q=100$, from the final configuration.}\label{fig:plotqual}
\end{figure}

The easiest way to get an idea of how the distribution of positions looks is to plot the values of the radius in the $12$ plane,  $r_{12}$ in various formats. For convenience, we also plot the spherical radius of the other four dimensions in $\BR^6$, which we call $r_{3456}$. An analysis of one configuration shows the following distributions, in the qualitative figure \ref{fig:plotqual}.

Analyzing the second figure carefully, we notice that the inner radius of the distribution of particles in $r$ has  a sharp edge to the left of a sharp peak, and is finished with a small smooth tail. This tail can probably be attributed to $1/N$ corrections and it might be possible to define the radius of the hole as  the locus of highest density. This is certainly geometric in character, but it depends on understanding the distribution of particles in the ensemble better. 
  If we were to plot instead the number of points in bins divided by the radius, as would be associated to a rotationally invariant measure $r_{12} dr_{12}$, the plot does not change significantly. The peak is enhanced a bit, and it moves a bit to the left. A definition of the radius of the hole based on this feature is interesting, but it is hard to understand how it would approach the large $N$ limit without extensive modeling.  Since we do not have an analytic understanding of these distributions yet, it is very hard to use their prominent features in a meaningful way to obtain a reasonable definition of the geometry. For illustration,  should we use the density in six dimensions to define the radius? Or should we use the projection of the density distribution to the $12$ plane? The choice would depend on which observable we need to evaluate and how it is realized geometrically in the ten dimensional dual geometry. Notice also that the definitions of the radius that we use in most cases do not suppress the peak very much relative to the inner radius. Because there are so many more particles there, one can expect that the radius measurements we are doing at $k=14$ are tracking the peak in the distribution above. The distributions above were plotted after the choice was made of which functions to average. 
  
  In hindsight, other definitions of the radius might have been better choices. After all, we actually know how the distributions were generated, so there is in principle a lot more information encoded in the eigenvalue distributions. In a certain sense, we are purposefully ignoring this information except for some qualitative understanding. The main reason to do this is to think of the algorithm that generated the geometry as a black box: the passage from wave functions to geometries is not really understood, so how the details of the wave function appear in the data are still obscured.  Given enough data, one should be able --at lest  in principle-- to reconstruct the `theory' from the output of the black box. What we have chosen to do is to use simple geometric observables that describe integrals over a geometry. If we had a more precise understanding of the data, this could be exploited to get a picture of local geometric information.
   
The next step in our procedure is to examine the data of the moments of the inverse radius of the distribution and to check how big are the fluctuations, relative to the size of the moments for the same setup $N=10000$ and $Q=100$. This is the choice that we made before looking at the data. In particular, we are at first more interested in seeing how large are the fluctuations of these observables. This will give us a rough idea of how far from the large $N$ limit we really are. In a certain sense, this is one of the most important pieces of information we can gather. If we expect some precision on the local geometry, we want basic geometric quantities to have low variance. If the variance is high, individual pictures of the data will be very fuzzy relative to the large $N$ limit.

\begin{figure}[ht]
\epsfxsize=6cm \epsfbox{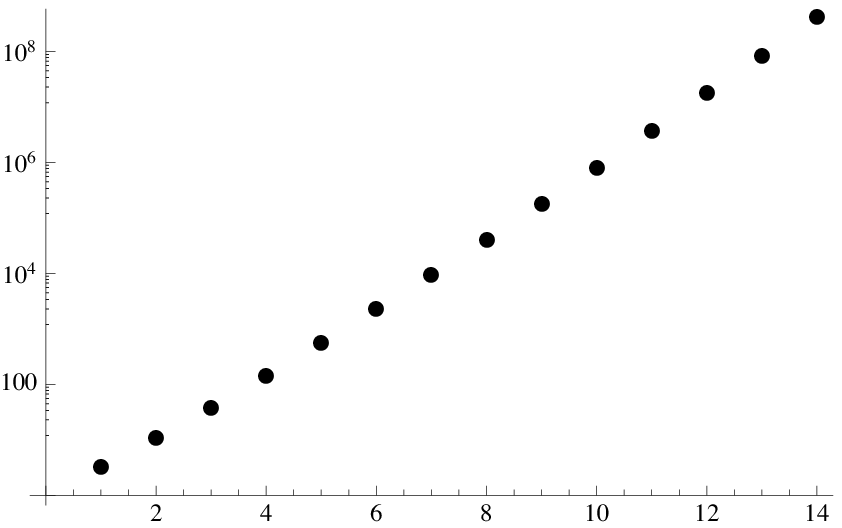}\epsfxsize=6cm\epsfbox{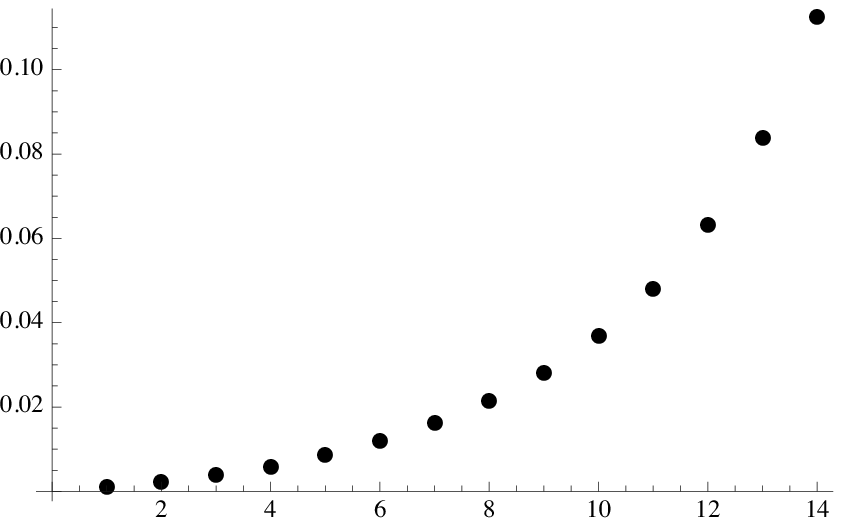}
\caption{On the left figure, the mean value of the moments at $N=10000$, $Q=100$, for $k=1,\dots, 14$. On the right,  the statistical variance  $\delta f_k/f_k$. As can be seen, the 
variance in the values increase as we take $k$ larger, and are already of order $10\%$ for $k=14$, which is a moderate value of $k$.  This indicates that fluctuations start becoming important for relatively low values of $k$, even in the case where $Q$ is large. For smaller $Q$ the inner radius is smaller, and fluctuations of one outlying particle near the inner edge are even more important in $f_k$.  }\label{fig:errorgrowth}
\end{figure}

The results of this analysis appear in the figure \ref{fig:errorgrowth}.

\begin{figure}[ht]
\epsfxsize=8cm \epsfbox{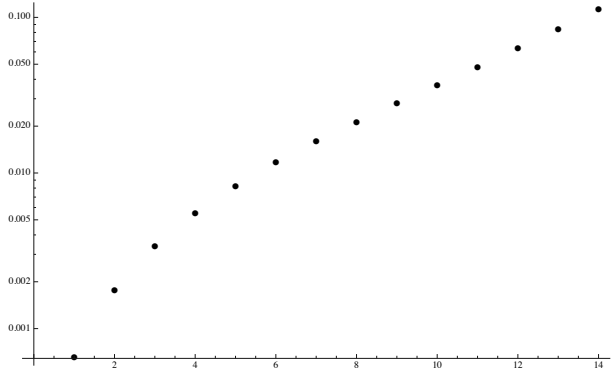}
\caption{A log plot of the data in figure \ref{fig:errorgrowth}}\label{fig:logerrorgrowth}
\end{figure}

The figure shows that the variance of the observables depends on the quantum numbers of the observables. A look of the data suggests a power law behavior. We can test this in a log plot, shown in figure \ref{fig:logerrorgrowth}.

 A fit to log suggests that $\delta f_k/f_k \simeq A k^{0.35}$, but looking carefully at the data in figure \ref{fig:logerrorgrowth} the power law behavior sets in after $k=5$ and for low values of $k$ the answer is different.
Since we do not have a theory of these fluctuations, we can not do a fit that is sufficiently informed to know if this behavior is typical and represents a departure from the large $N$ limit or not. Also, one show notice that since all of the variables we are measuring are strictly positive, the distribution of the variables is not gaussian. For $\delta f_k/f_k$ sufficiently small this should not matter too much. However for large $k$ this is important and therefore the measurement of the fluctuations will have (large) non-gaussianities. Without a statistical model for how to deal with these issues, the numbers above just give order of magnitude estimates for deviations from typicality. 
 
\begin{figure}[ht]
\epsfxsize=8cm \epsfbox{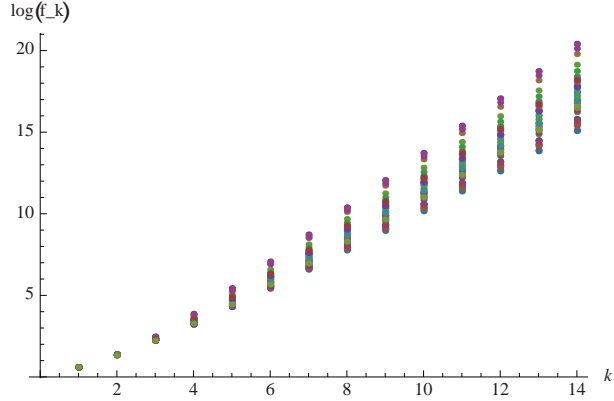}
\caption{A plot of $log_{10}(f_k)$ versus $k$ for the case of $N=1000$, $Q=5$ for $50$ different configurations. Notice how for large $k$ the magnitudes cover various decades. }
\label{fig:hugefluctuations}
\end{figure}

Now, we can go to a more extreme case, where the $1/N$ corrections are much larger. We do this in the figure \ref{fig:hugefluctuations}, which is worked for the case $N=1000, Q=5$. Notice that in this case the sizes of the different $f_k$ changes by various orders of magnitude at high $k$ between different configurations. Hence the fluctuations are very large and dominate any calculation. This is a situation where we expect that the averages $\vev{f_k}$ do not converge for $k$ large, as described in the previous section. Hence the behavior of the graph is as would be expected: the large values of $f_k$ are dominated by large fluctuations rather than by the mean values. 

\begin{figure}[ht]
\epsfxsize=10cm \epsfbox{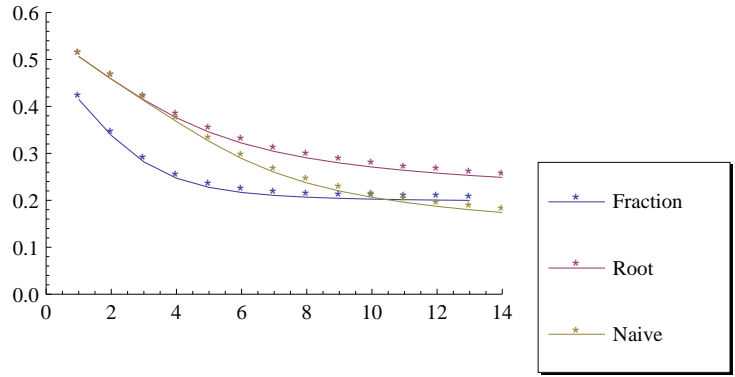}
\caption{The radius  $R_{0}$, $r^{(r)}$, $r^{(s)}$ of the hole relative to the radius of the sphere at $Q=0$, evaluated at different values of $k$ before taking the limit . We see that the values of $R_0$ are lower than $r^{(r)}$ and show faster convergence,  while $r^{(s)}$, the naive definition of the radius converges towards zero and is dominated by fluctuations for large $k$. }
\label{fig:compare}
\end{figure}

For the same extreme setup, we can also plot the radius as measured by the three different measurements that we suggested in the previous section. This appears in the figure \ref{fig:compare}. As can be read from the figure, we find that 
for various $k$, the definition of the radius $R_0$ converges faster than the definition of the radius based on the roots of 
$f_k$. We can prove this in general (the argument is found below). Moreover, we see that the naive definition of the inner radius crosses the one defined by the fractions, even though configuration by configuration the naive definition gives a higher value. This is because the averaging procedure is different. Notice also how for low values of $k$ the naive definition and the root definition agree for the most part. This is exactly what is expected when fluctuations are small.

Now, let us prove that the inequalities $R_{0k}<r_{k+1}^{(r)}$. Notice that configuration by configuration we have the following  equality 
\begin{equation}
\frac{f_{k}}{f_{k+1}}= \frac{r_k^{-2k}}{r_{k+1}^{-2(k+1)}}= \left(\frac{r_k}{r_{k+1}}\right)^{-2k} r_{k+1}^2
\end{equation} 
We also have the the inequality given by
\begin{equation}
r_{k+1}< r_{k}
\end{equation}
This is obvious, as the $r_{k+1}$ is dominated more strongly by the particles near the inner edge than $r_k$. It can also be proved mathematically using H\"olders inequality or Jensen's inequality.

From here, it follows that 
\begin{equation}
\sqrt{\frac{f_{k}}{f_{k+1}}}< r_{k+1}
\end{equation}
And this inequality persists when we take averages. Hence the claim. Also, in the limit $k\to \infty$, both definitions would converge to the same value (before averaging the limit is exclusively dominated by the innermost particle), but the one based on the ratios converges faster (what we called $R_0$). Hence the definition $R_0$ is  the most numerically economic calculation of the radius of the configuration, as promised earlier. For completeness, we show now the plot in figure \ref{fig:compare} with the variance of the data set included.

\begin{figure}[ht]
\epsfxsize=8cm\epsfbox{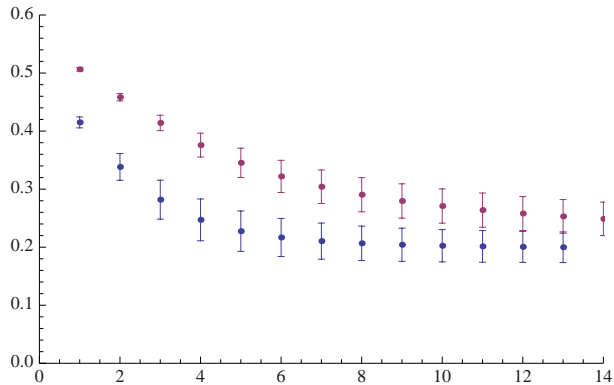}
\caption{Plot of $R_{0k}$ and $r_k^{(r)}$ with the variance of sample included: this would indicate the error of a single measurement on the quantum system. It essentially accounts for the quantum fluctuations. This is the same data set used to generate figure \ref{fig:compare}.}
\end{figure}

As should be recognized from the analysis above, the choice of variable to measure a geometric property in a model of emergent geometry with quantum fluctuations can receive very large corrections due to the fluctuations. This depends on how we average over configurations. This is determined in essence by the observable that one chooses to probe the system with. Although this is not surprising, the fact that fluctuations can be rather large even for large Values of $N$ ($N=10000$ in our example) show that even though they are nominally suppressed by $1/N^2$, the coefficient in front of this suppression can be very large depending on the `quantum numbers' of the observable. This suggests that for many detailed geometric questions we might need to consider very large values of $N$. For simulations on computers, this requires a lot of resources and the development of fast algorithms to perform the task efficiently.

\subsection{The approach to large $N$: fixed $q$ simulations.}

\begin{figure}[ht]
\epsfxsize=10cm\epsfbox{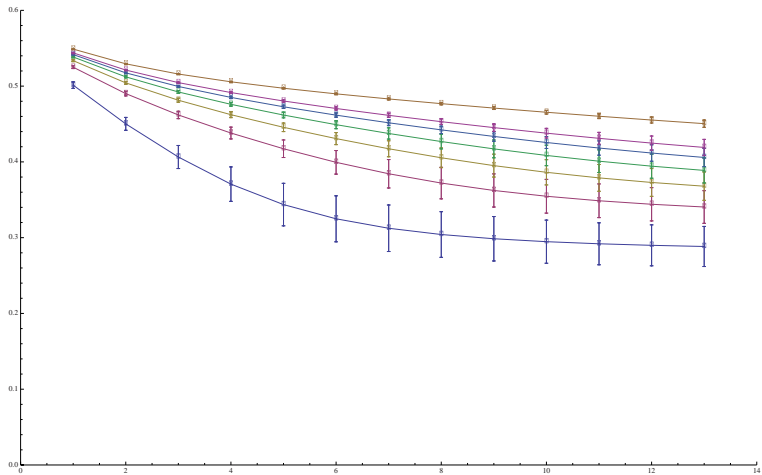}
\caption{The approach to large $N$.  We show data for $R_0$ for different values of $k$, at $q=0.01$ and $N=1000,2000,3000,4000,5000,6000,10000$ from bottom to top in the graph. The error bars signify the variance
of the measurements. Again, this signifies the size of the quantum fluctuations.}
\label{fig:largeNapp}
\end{figure}

The next step in our analysis is to check how the results converge to large $N$ values for objects that are comparable. In this case, we need to keep $q=Q/N$ fixed to match the thermodynamic limit. We choose $q=0.1$ and plot our results for various values of $N$. This is shown in figure \ref{fig:largeNapp}. As can be seen for the figure, as $N$ gets larger, the reported value of the radius of the configuration increases. This is expected for the simple reason that at smaller $N$ fluctuations of particles near the inner radius are bigger, and the observables we have are sensitive to these fluctuations.
Notice also that the fluctuations in individual measurements as a function of $N$ decay with $N$. However, from the data with the smallest value of $N$ to the highest value of $N$ we have one decade in difference, and that seems to be the same difference in the size of the fluctuations.

We can also see that the differences between the observables calculated at different $N$ that are separated by the same amount are getting smaller as we increase $N$. This strongly suggests that there is convergence to a large $N$ limit, but graphically we see that the convergence is rather slow.

 This suggests that the deviations go as $1/N$ rather than $N^{-2}$. This is rather surprising. This result should be contrasted with  the results \cite{Yaffe}, where the limit $N\to \infty$ should be a classical system and we should be able to learn everything about the system expanding in powers of $1/N^2$. 

However, in a sense, the model is not a matrix model in the t' Hooft limit \cite{'tH}. We have taken the t' Hooft coupling $\lambda= g^2N\to \infty$ keeping $N$ fixed and large, so that we can reduce our model to the set of commuting matrices. But a similar step of expanding about commuting matrices was done in \cite{BHH} for a non-supersymmetric model. There it was found that the joint eigenvalue distribution obtained by taking this procedure was not exactly the correct size and that ignoring the loops of the variables that were integrated out gave the wrong answer. If we assume that something similar is happening here, where supersymmetry makes cancellations between integrated out modes happen, then we might still find problems if cancellations are not exact. This might be more pronounced at the edges of the distributions where small corrections have the biggest effects. At this stage, we are just speculating on what might be the cause for this feature of the numerical simulations. Another issue relates to the observations in the figure \ref{fig:hugefluctuations} which show that for large $k$ we are not in the factorization regime yet for small $N$. Thus the idea of seeing a natural convergence to a standard large $N$ expansion is suspect because $N$ is not large enough yet. This is a general issue that one should be very cautions about before making claims of some effect. Different observables converge at different rates,  and the corrections depend
on the parameters describing the observable. For example, if in our setup above corrections go like $k^s/N$, and $s$ is large, then we need to go to very large $N$ before making any claims, because for $k=10$ the result could easily overwhelm $N$. Either way, for certain questions, a very large value of $N$ will be required before being able to compare to $N=\infty$. Considering also that the total number of degrees of freedom in  the system before gauging is $N^2+5N$ and that the 
`energy of the ground state' is of order $N^2$ plus corrections of order $N$, it is perhaps not surprising that 
the naive planar counting argument \cite{'tH} fails to give the right approach to the large $N$ limit. After all, we are taking  the planar parameter to infinity first (we are integrating out heavy modes and assuming they have no corrections to the truncated theory), and we keep $N$ finite. Still, this is a puzzle that we can not resolve satisfactorily yet.
Further study is required to understand this issue better.

With our data, we can  find a fit to the exponent and see what happens. We can do the following analysis.  We can assume that there is a limit and a scaling exponent that describes how close we are to the limit, by the following argument
\begin{equation}
S_N \simeq S_{\infty} + \frac{\Delta S }{N^\alpha}+\dots
\end{equation}
Then, if we take the following differences 
\begin{equation}
S_N-S_{2N}= (1-2^{-\alpha}) \frac {\delta S}{N^\alpha}
\end{equation}
we can make a fit for $\alpha$ and check the exponent appearing in the corrections. The data is not very illuminating. We find values as large as $\alpha\sim 0.7$ for low values of $k$ up to $k=6$, and then it decreases towards $\alpha\sim 0.1$ for  $k=13$. Again, we should not give too much meaning to these numbers for the large values of $k$ if the corresponding correlators are not factorized yet.

\begin{figure}[ht]
\epsfxsize=8cm\epsfbox{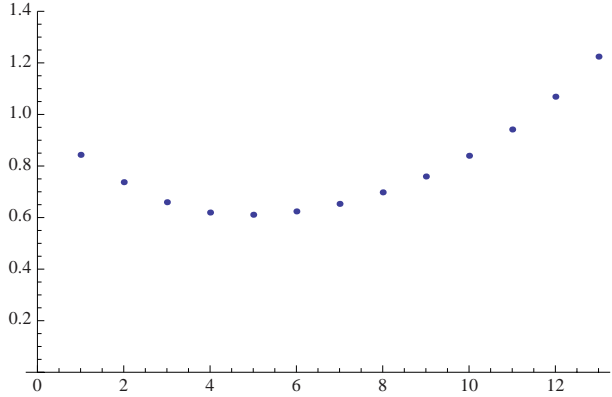}
\caption{Scaling exponent for the variance of $R_0$ at various values of $k$. } 
\label{fig:fluctscaling}
\end{figure}

Similarly, the standard-deviation fluctuations should have some scaling pattern 
\begin{equation}
\delta S_N \simeq \frac{A_N}{N^\beta}
\end{equation}
If we plot the data sample to the best fit for $\beta$ we get the results in figure \ref{fig:fluctscaling}. The results are roughly consistent with $\beta\sim 1$, which is the exponent that the planar counting would suggest, but it is unclear how this data should be used, seeing as we might not yet have converged to large $N$ sufficiently.  

Again, rough consistency with $\beta \sim 1$ is not the same as knowing that $\beta=1$. A much larger data set from more simulations would be able to verify this hypothesis. It would also require going to much large values of $N$ (hopefully stepping up to a factor of 100 in $N$ might be enough).  

If the planar counting does not work, it could be because we ignored some degrees of freedom that contribute to the limit, and that these make the full system behave as is expected.
 However, we should expect that the corrections induced by integrating out heavy modes involve their masses, which are expressed as powers of the large $N$ coupling constant $\lambda=g^2N$ \cite{BLargeN,BCV}.
 We do not have a clear understanding on how the planar counting would be restored once we take $\lambda$ finite. It would require subtle cancellations to take place in order to do that. The upshot is that this is an interesting observation that the numerical exploration is giving us. It would be very hard to guess this behavior for observables with our current understanding of large $N$ systems \cite{Yaffe}. The one thing that is very special of this setup is that when we look at the eigenvalue density distribution in the saddle point limit, it is  singular: it has a $\delta$ function behavior. 
 Singular distributions get smeared by fluctuations and because they are singular, they might induce anomalous scalings of fluctuations, etc. Again, a thorough understanding of these issues requires further work.

\subsection{A scaling relation: size of the hole for varying $q$ at fixed $N$}

\begin{figure}[ht]
\epsfxsize=8cm\epsfbox{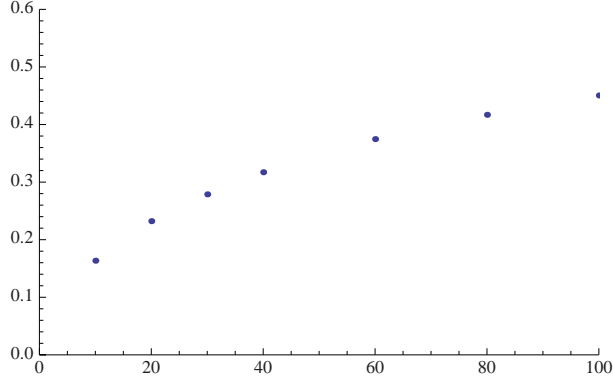}
\caption{Plot showing Dependence on the size of hole to the parameter $Q$ , at $N=10000$}
\label{fig:qdepend}
\end{figure}

The next item in the list is to check some of the geometric features of our ensemble and compare them to the expected solutions of supergravity and in particular with the free fermion picture \cite{Btoy,LLM}. The idea is to fix $N$ and vary $Q$, thereby exploring the parameter $q=Q/N$. We do this for our largest value of $N=10000$ and for the largest $k$ that we recorded $k=13$. We plot $R_0(q)$ versus $Q$ in the figure \ref{fig:qdepend}.

\begin{figure}[ht]
\epsfxsize=8 cm \epsfbox{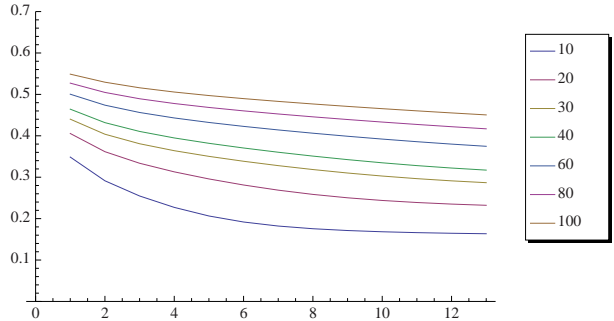}
\caption{The function $R_0 (Q)$ at various values of $k$, for $Q=10,20,30,40,60,80,100$, $N=10000$}
\label{fig:qdepall}
\end{figure}

A fit of the form $R_0(q) \sim q^\alpha$ shows that the best fit is with $\alpha\simeq 0.47\pm 0.1$, where the error is  an  estimate of the systematic error. We use the variance in the inner radius as a measurement of the systematic uncertainty. This is consistent with $\alpha=0.5$. Incidentally, $\alpha=0.5$ is the scaling value for the free fermion droplets. The reason for this is that the free fermion system (quantum hall system) is equivalent to a Coulomb gas in two dimensions. For that system both particles and holes form incompressible fluids, and the radius of the inner hole in these systems also scales like $R\simeq Q^{0.5}$ where $Q$ is the number of holes. 
The full data for all $k$ for the same data set is shown in the figure \ref{fig:qdepall}. This can be used to estimate errors. As can be seen, for large $Q$ the data has not converged yet, so the values quoted in the figure \ref{fig:qdepend} are high and taking care of this more systematically  has a tendency to reduce the exponent $\alpha$. It is hard to extrapolate to large $k$, because we do not have an exact functional form to match. This requires understanding the statistics and the large $N$ limit better. As seen in the previous section, this is rather subtle.
  
In a previous paper \cite{BCotta}, a different measurement was done by keeping $Q$ fixed but varying $N$, and a scaling exponent of $1/4$ was reported. This expression involved small values of $N$, so it is very likely that the system was subject to large fluctuations. 
 In this paper we have seen how this is an issue in general, and the deviations can be very large. Also, the convergence to large $N$ is slow, so the dependence on $q=Q/N$ is not reliable in the paper \cite{BCV}. The range of $q=Q/N$ explored in that paper was similar to the one of this paper. 
Neither result is conclusive. In this paper the radius has not converged and the systematic errors of extrapolation make the determination of $\alpha$ done here somewhat unreliable. The best way to reduce errors is to have larger simulations and to improve the understanding of the limit $k\to \infty$. What we see is that $k\simeq 14$ is still too low. We did not know this ahead of the simulation and unfortunately larger values of $k$ were not recorded. This decision was  made ahead of the simulations: we only stored moments of distributions rather than full configurations. Full configurations generate very large data files in order to get large enough statistics. Due to various limitations, we opted for this approach.
In a sense, we have performed this analysis as a blind analysis on the data. Consider however, that the information 
we have gathered is important from another perspective. If one sets out to do large simulations, the problems of data management is an issue. Our exploration gives us some rather precise information of how much data we need to keep and
in what type of format. 
 
As a further thought, one expects that there are corrections to the scaling behavior that would involve some power series of $q$. Such corrections would require much more data to do a proper fit, as well as a systematic understanding of the functional form of the correction. If this is studied,  the dominant scaling exponent (that with the smallest power of $\alpha$) would be revealed for $q$ small. 

It would certainly be interesting to study these puzzles further and to match some properties to the gravity theory in observables that can be matched on both sides with better understanding (for example the energies of string states, as was done for the ground state in \cite{BCV}).

\section{Conclusion}

In this paper we have examined the problem of extracting geometric data from a system that exhibits emergent geometry in a thermodynamic limit where the number of particles $N$ in a statistical mechanical system becomes large. The geometric data we computed was one single parameter: the radius of a feature, that depends functionally on  a variable defining a one parameter family of wave functions of the system. 

In order to measure this geometric feature,  we needed some scheme to define the shape parameter at finite $N$. We chose observables that were simple to calculate numerically and that have a reasonable behavior at large $N$ as given by our theoretical expectations of what constitutes a reasonably good observable.
 Our work has shown that extracting the geometric data requires making judicious choices of these observables. Otherwise, one might find pathological distributions whose averages are infinity entering into the problem when $N$ becomes finite, even though when $N$ is very large they seem to be well behaved. Learning how to avoid pathological measurements requires some trial and error and we gave some examples of how to improve this simple class of observables so that they had better behavior.
 
We also studied how large were the fluctuations on the observables we chose and how fast the results approached the large $N$ limit.
From the data we could analyze, the corrections to large $N$ are controlled by various powers of $1/N$, and some of them seem to be inconsistent with the naive planar counting lore. Surprisingly, the convergence to large $N$ was slower than expected, and the fact that some observables had pathologically large fluctuations renders a lot of the analysis inconclusive: even with $N\simeq 10000$, $N$ was not large enough to see a clear pattern emerging from the simulations. From this point of view we have learned a lot about the scale of simulations that would be required to sort out these features with the same procedures in future studies. We should have much larger simulations, with $N\simeq 10^6$.

Part of the reason why such large numbers are required should be obvious to anyone that tries to understand higher dimensional geometries of dimension $d$ by adding a discretized grid approximation to the geometry. If we want to double the spatial resolution, we need a factor of $2^d$ enhancement on the number of cells. As $d$ becomes larger, this is an issue. A value of $N\simeq 10^4$ is small to refine five dimensional geometries by these standards.

Clearly, any future study should require smarter and faster algorithms than what we have used so far. We have used a simple Monte Carlo that runs on a single processor for these simulations.  A better algorithm would involve parallel processing, a faster procedure of generating data sets and it might also require some coarse graining improvements to perform calculations with a study of the consequent systematic errors introduced by approximations. 

Another feature that we discovered is that it is very likely that there are better ways of analyzing the data than what we did. Particularly, features of the data as shown in figure \ref{fig:plotqual} suggest that that the density profile in the various variables is a rich area to explore. If we improve our theoretical understanding of the data this might be exploited to address issues of geometry further. For example, in our analysis the fact that the geometry is a submanifold of $\BR^6$  was not crucial. However, it is a robust theoretical feature that should be exploited to analyze the system better. 

For geometries with more features than a single hole in the center of the distribution, it would be interesting to figure out if one can find a fit to the wave function from numerical data. Such a dictionary would help us understand a lot better the
relation between the geometry and  the original quantum setup. Such an approach will obviously use more information about how the configurations were generated to infer (or reconstruct) information.

Notice also that we are not considering either coarse grained geometries (statistics over multiple states that share common features) nor random statistical states in the ensemble. For such states, other techniques are available that show how hard it is to measure microscopic geometries \cite{BCLS}. Distinguishing the geometries obtained from bona-fide 'geometries' from those obtained from random superpositions of states should also be interesting. 
We are currently looking into these matters.

\acknowledgments

D. B. would like to thank G. Horowitz and especially L. Yaffe for many discussions on large N limits. Work supported in part by DOE under grant DE-FG02-91ER40618.

\end{document}